\documentclass[fleqn,usenatbib]{mnras}

\usepackage{newtxtext,newtxmath}
\usepackage{amsmath,amssymb, graphicx,xcolor}
\usepackage[export]{adjustbox}

\usepackage[utf8]{inputenc}

\usepackage[compact]{titlesec}  
\titlespacing{\section}{0pt}{12pt}{7pt}
\titlespacing{\subsection}{0pt}{9pt}{4pt}

\newcommand{\frb}{FRB~180916.J0158+65}
\newcommand{\PRESTO}{\texttt{PRESTO}}
\newcommand{\flu}{\,Jy\,ms}

\usepackage[T1]{fontenc}

\DeclareRobustCommand{\VAN}[3]{#2}
\let\VANthebibliography\thebibliography
\def\thebibliography{\DeclareRobustCommand{\VAN}[3]{##3}\VANthebibliography}

\usepackage{graphicx}	
\usepackage{amsmath}	
\usepackage{amssymb}	

\title[\frb\ with the uGMRT]{Detection of 15 bursts from \frb\ with the uGMRT}

\author[Marthi et al.]{V.~R.~Marthi$^1\thanks{Email: \href{mailto:vrmarthi@ncra.tifr.res.in}{vrmarthi@ncra.tifr.res.in} (VRM); \href{mailto:dzli@cita.utoronto.ca}{dzli@cita.utoronto.ca} (DZL); \href{mailto:ramain@mpifr-bonn.mpg.de}{ramain@mpifr-bonn.mpg.de} (RAM); \href{mailto:wharton@mpifr-bonn.mpg.de}{wharton@mpifr-bonn.mpg.de} (RSW)}$, T.~Gautam$^2$, D.~Z.~Li$^{3,4,2}\textcolor{blue}{\footnotemark[1]}$, H-H.~Lin$^{3,2}$, R.~A.~Main$^2\textcolor{blue}{\footnotemark[1]}$, A.~Naidu$^{5,6}$,
U-L.~Pen$^{3,7,8,9,2}$ and \newauthor R.~S.~Wharton$^2\textcolor{blue}{\footnotemark[1]}$\\
$^{1}$National Centre for Radio Astrophysics, Tata Institute of Fundamental Research, Post Bag 3, Ganeshkhind, Pune - 411 007, India \\
$^{2}$Max-Planck-Institut f{\"u}r Radioastronomie, Auf dem H{\"u}gel 69, D-53121 Bonn, Germany \\
$^{3}$Canadian Institute for Theoretical Astrophysics, 60 St. George St. Toronto, Canada \\
$^{4}$Department of Physics, University of Toronto, 60 St. George Street, Toronto, ON M5S 1A7, Canada\\
$^{5}$Department of Physics, McGill University, 3600 rue University, Montréal, QC H3A 2T8, Canada\\
$^{6}$Oxford Astrophysics, Denys Wilkinson Building, Keble Road, Oxford, OX1 3RH, United Kingdom\\
$^7$Dunlap Institute for Astronomy and Astrophysics, University of Toronto, 50 Saint George Street, Toronto, ON M5S 3H4, Canada\\
$^8$Program in Cosmology and Gravitation, Canadian Institute for Advanced Research, Toronto, ON M5G 1Z8, Canada\\
$^9$Perimeter Institute for Theoretical Physics, 31 Caroline Street North, Waterloo, ON N2L 2Y5, Canada
}

\date{Accepted XXX. Received YYY; in original form ZZZ}

\pubyear{2020}

\begin{document}
\label{firstpage}
\pagerange{\pageref{firstpage}--\pageref{lastpage}}
\maketitle

\begin{abstract}
We report the findings of a uGMRT observing campaign on \frb, discovered recently to show a 16.35-day periodicity of its active cycle.  We observed the source at 550-750\,MHz for $\sim 2\,$hours each during three successive cycles at the peak of its expected active period. We find 0, 12, and 3 bursts respectively, implying a highly variable bursting rate even within the active phase. We consistently detect faint bursts with spectral energies only an order of magnitude higher than the Galactic burst source SGR~1935+2154. The times of arrival of the detected bursts rule out many possible aliased solutions, strengthening the findings of the 16.35-day periodicity. A short-timescale periodicity search returned no highly significant candidates. 
Two of the beamformer-detected bursts were bright enough to be clearly detected in the imaging data, achieving sub-arcsecond localization, and proving as a proof-of-concept for FRB imaging with the GMRT. We provide a $3\sigma$ upper limit of the persistent radio flux density at 650\,MHz of $66~\mu{\rm Jy}$ which, combined with the EVN and VLA limits at 1.6~GHz, further constrains any potential radio counterpart. These results demonstrate the power of uGMRT for targeted observations to detect and localize known repeating FRBs.
\end{abstract}




\section{INTRODUCTION}

Since their discovery more than a decade ago \citep{lorimer2007}, Fast Radio Bursts (FRBs) have excited considerable interest in their origin, which remains yet unknown. The observed dispersion measures (DMs) of these bursts far exceed the expected contribution from the Milky Way, suggesting that FRBs are located at cosmological  distances. Indeeed, sub-arcsecond localizations for five of the $\sim$100 known FRBs have placed them in host galaxies with redshifts $z = 0.03-0.66$ (\citealt{chatterjee2017}, \citealt{bannister2019}, \citealt{ravi2019}, \citealt{prochaska2019}, \citealt{mnh+20}). A small subset of FRBs are seen to repeat, providing an excellent laboratory for understanding the physical mechanism that produces FRBs since we can conduct targeted observations that are not possible with any other FRB source.

\frb\ is the most active repeating source detected at 400-800 MHz 
by the Canadian Hydrogen Intensity Mapping Experiment 
\citep[CHIME,][]{chime2019b}, with 52 bursts reported as of 20 July 2020\footnote{\url{https://www.chime-frb.ca/repeaters}}.
Recently, the European VLBI Network (EVN) localized it to a star-forming region in a spiral galaxy at $z = 0.0337$ ($D_{\rm L}\approx 150$~Mpc, \citealt{mnh+20}). This is by far the closest localized extragalactic FRB source, an order of magnitude closer than other localized sources. 

Remarkably, a periodicity in the activity of \frb\ has recently been discovered. All CHIME-detected bursts occur within a 5-day window that repeats every 16.35 days \citep{chime2020a}. This is the first detected periodicity in a repeating FRB and presents a unique opportunity to unveil the nature of FRB sources. The periodicity could be caused by the orbital motion of the FRB source around a companion (\citealp{16DaiOrbit}; \citealp*{lyutikov+20,20IokaOrbit}), 
precession of the burst emitting object itself \citep*{20ZanazziPrecession,levin+2020,20YangPrecession} or ultralong rotational periods \citep*{Beniamini2020a}. \citet{20Rajwade} have similarly detected a tentative $\sim$157d periodicity in FRB 121102, an order of magnitude larger. 
The ability to predict the next active phase of the repeating \frb\ means that highly efficient follow-up observations can be conducted with sensitive radio telescopes. Regular monitoring observations can hence be crucial in pinning down the FRB emission mechanism, and  especially in bridging the energy gap with Galactic radio transient counterparts. As well, they can help us understand the origin of its periodicity. 

Here we present the findings from observations of \frb\ during three successive active windows taken as part of an ongoing monitoring 
campaign with the upgraded Giant Metrewave Radio Telescope (uGMRT; \citealt{uGMRTpaper}). This letter is organized as follows.  In Section~\ref{sec:observations} we describe our observations, data reduction, and burst search methods.  Section~\ref{sec:results} details our results, including our detected bursts, a short-timescale periodicity search, resolving aliasing ambiguities in the 16.35-day periodicity, burst localizations and limits on the flux of the persistent source. Finally, Section~\ref{sec:discussion} contains our concluding remarks.

\section{OBSERVATIONS}\label{sec:observations}
The uGMRT observations were carried out on 2020 March 09, 2020 March 24, and on 2020 June 30, all near the peak of the active phase within the 16.35-day period. We used the uGMRT Band-4 receiver at 550-750 MHz, a subset of the CHIME band.  The GMRT Wideband Backend (GWB; \citealt{reddy+17}) was deployed in the 200-MHz 8-bit beamformer mode. Interferometric visibilities from the same FX correlator were simultaneously recorded. In the beamformer mode of operation, the GWB is capable of computing four simultaneous independent beams. For our observations, we recorded the coherently dedispersed (CD) phased array beam. The correlator was set up to split the 200-MHz bandwidth into 2048 channels. The beam output was integrated to 327.68 $\mu$s, while the visibilities were integrated to 2.68 s.

In each session we chose 0217+738, an unresolved radio source $\sim 10^\circ$ away from our target, as the phase calibrator. We have a total of 110 minutes of exposure in each of the first two sessions and 85 minutes in the third.

\subsection{Detecting the bursts}
\begin{figure*}
    \includegraphics[width=1.0\linewidth,scale=1.0,trim=2.0cm 8.0cm 3.0cm 2.5cm, clip=true]{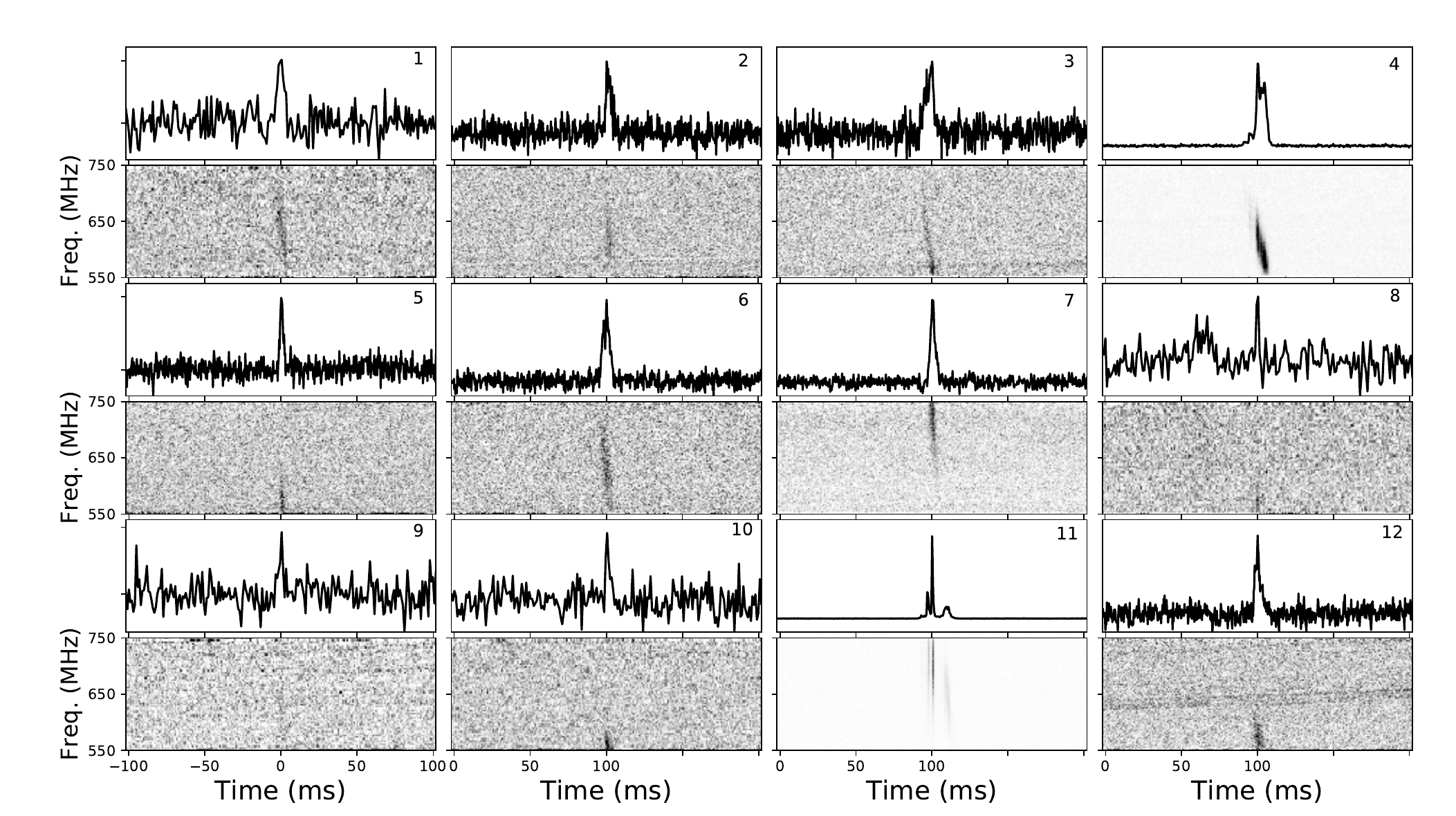} \\
    \vspace{2mm}
    \includegraphics[width=1.0\linewidth,scale=1.0, trim=2.0cm 2.6cm 3.0cm 21.0cm, clip=true]{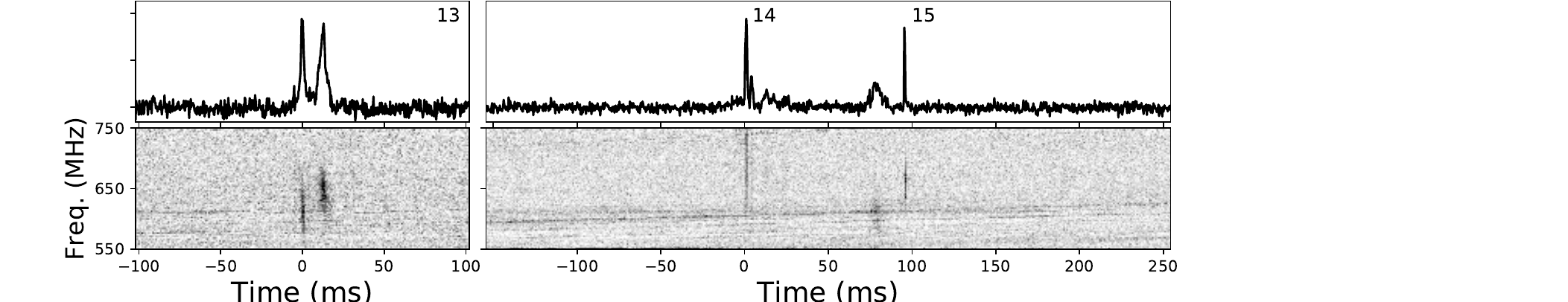}
    \vspace{-4mm}
    \caption{Dynamic spectra of all detected bursts, in sequential order.  For plotting purposes, frequency was binned by a factor of 16.  
    In bursts 1, 8, 9 and 10, frequency was binned by an additional factor of 2, and time by a factor of 3 to $983.04\,\mu$s.
    The top panels show our detections from March 24, while the bottom row are from June 30, where the final panel is interpreted as at least two separate bursts. }
    \label{fig:burstpanorama}
\end{figure*}

\begin{figure}
    \includegraphics[width=0.541\linewidth,scale=1.0, trim=0.0cm 0.0cm 0.0cm 0.0cm, clip=true]{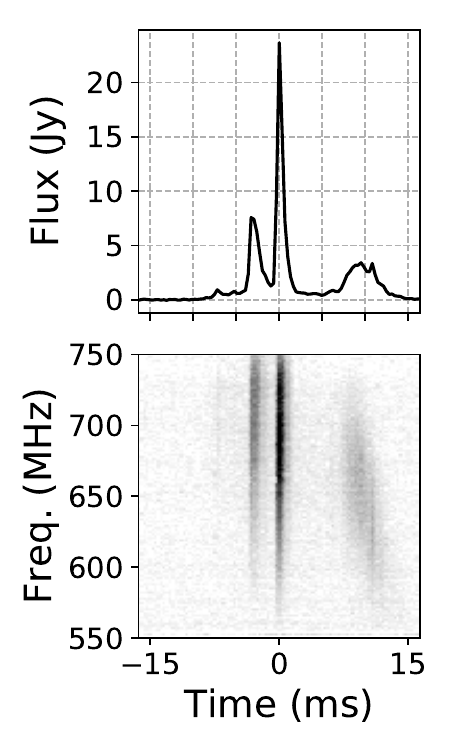}
    \includegraphics[width=0.45\linewidth,scale=1.0, trim=0.5cm 0.0cm 0.0cm 0.0cm, clip=True]{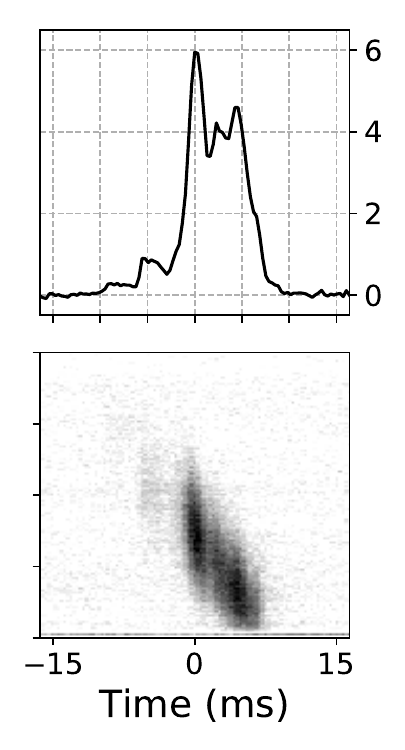} \\
    \vspace{-3mm}
    \caption{Zoomed in dynamic spectra and profiles of burst 11 (\textit{left}), and burst 04 (\textit{right}).  The grayscale scheme is in square root scaling to reveal the faint emission.  While both bursts show clear sub-structure and multiple bright peaks, there is persistent detectable emission for the full $\sim 20$\,ms duration of the bursts. }
    \label{fig:dynspec-brightest}
\end{figure}

We performed a standard \PRESTO\ \citep*{Ransom2002}
search on the coherently dedispersed phased array data. We flagged RFI in 2\,s windows using the \texttt{rfifind} algorithm, and used the \texttt{prepsubband} command to create dedispersed timeseries in 201 steps of $0.5$\,pc cm$^{-3}$, with the DM centered at 348.82\,pc cm$^{-3}$. We searched for bursts using  \texttt{single\_pulse\_search.py}, identifying candidates above 6$\sigma$, distinguishing candidates from RFI by eye. On 2020 March 24, 2020 June 30 and 2020 March 09, 12, 3 and no bursts were detected respectively, shown in Figure~\ref{fig:burstpanorama}. The profiles and dynamic spectra of bursts 11 and 04, the brightest pair among those reported here, are shown in Figure~\ref{fig:dynspec-brightest}. The time of arrival (ToA) for each burst was obtained as the offset from the barycentered start time of the respective scan, returned by \PRESTO.

\begin{table}
    \centering
    \begin{tabular}{|r|c|r|r|r|}
    \hline
    Burst  & Barycentric   & Peak  & Fluence  & Error    \\
          &         TOA   & flux  &          &          \\
     \#    &   MJD58932+    &   Jy      &  \flu   &  \flu\\
    \hline
     01   & 0.542126        &  0.19    &   0.26  &  0.12 \\
     02   & 0.542939        &  0.32    &   0.65  &  0.23 \\
     03   & 0.544481        &  0.27    &   0.37  &  0.15 \\
     04   & 0.550594        &  5.98    &  32.90  &  9.91 \\
     05   & 0.557890        &  0.38    &   0.24  &  0.10 \\
     06   & 0.563082        &  0.57    &   1.54  &  0.50 \\
     07   & 0.567436        &  0.83    &   6.68  &  2.04 \\
     08   & 0.579496        &  0.21    &   0.12  &  0.06 \\
     09   & 0.580353        &  0.17    &   0.09  &  0.06 \\
     10   & 0.588714        &  0.21    &   0.10  &  0.06 \\
     11   & 0.604069        & 23.65    &  47.80  & 14.39 \\
     12   & 0.612024        &  0.57    &   3.86  &  1.20 \\ 
      \hline
          &    MJD59030+       &          &         &    \\
      \hline
     13   & 0.297435         &  0.58    &   0.82  &  0.42 \\
     14   & 0.298753         &  0.80    &   6.63  &  0.74 \\
     15   & 0.298754         &  0.72    &   2.97  &  0.50 \\
     \hline
    \end{tabular}
    \caption{The ToA of the 15 bursts and their fluences. Only frequencies 569.5-745.4 MHz (channels 200 to 2000) were included for estimating the fluence.}
    \label{tab:fluences}
\end{table}

\subsection{Flux density calibration}
\label{section:fluence}

On 2020 March 24, we used our phase calibrator 0217+738, a flat spectrum 2~Jy source, also as the flux calibrator. We therefore allow for a modest 30\% systematic uncertainty in the derived system-equivalent flux density (SEFD), included in the error bar of the fluence for each burst in Table~\ref{tab:fluences}.

However on 2020 June 30, we observed 3C147, a bright flux calibrator. We therefore absorb only a 10\% systematic uncertainty in the SEFD, again included in the fluences of those bursts. We find that the SEFD on the two days, and therefore the peak flux densities of bursts with similar S/N in the two sessions, are comparable, inspiring confidence in our calibration of March 24 data.

The ToA, along with the identifier number, the peak flux density, fluence and the error bar on the fluence for each burst is given in Table~\ref{tab:fluences}

\section{RESULTS}
\label{sec:results}

\subsection{DM optimization with the power spectrum}
The burst profile obtained with a DM that maximizes the peak S/N is degenerate with any intrinsic temporal substructure. It is therefore more meaningful to maximize temporal substructure, recognizing its importance for investigating emission models and mechanisms.

We have developed a new method to determine the optimal DM that maximizes energy in the substructure, with similar motivation to the substructure maximization developed for FRB 121102 \citep{hessels+19}. First, we obtain the SVD of the time-frequency burst profile. Next, we identify the signal-free modes in the SVD, using which we reconstruct the noise and subtract it from the input burst profile. Finally, we produce the power spectrum of the noise-subtracted burst profile for a range of DMs around the nominal value, examining the power in the highest noise-free frequency mode as a function of the DM. The peak and the width of the Gaussian fit to the power yield the DM that maximizes substructure. For three selected bursts, namely 04, 11 and 12, we obtain optimized DM values of $349.0\pm 0.3$ pc~cm$^{-3}$, $348.8 \pm 0.1$ pc~cm$^{-3}$ and $348.9\pm0.9$ pc~cm$^{-3}$ respectively, consistent with the nominal value of $348.82$ pc~cm$^{-3}$ used throughout this work. The results of applying this technique to all the detected bursts will be detailed in an upcoming paper(Lin et al. 2020, \emph{in prep.}).

\subsection{Energy scales}
We detected bursts in a range of fluences between $\sim$ 0.1$-$48\flu, the lower end of this range being comparable with the faintest burst detected with the EVN of $0.2$\flu. Using the luminosity distance of 149.0$\pm0.9$\,Mpc \citep{mnh+20}, the spectral energy of our faintest detected burst (assuming an isotropic source) is $\sim 2.7\pm1.6 \times 10^{27}$ erg\,Hz$^{-1}$ 

Recently, the brightest known Galactic radio burst to date was detected by CHIME \citep{chime2020b} and STARE2 \citep{bochenek+20}, originating from the Galactic magnetar SGR 1935+2154. With an isotropic-equivalent spectral energy release of $1.6\pm0.3 \times 10^{26}$\,erg\,Hz$^{-1}$ it is approaching the energy scales of FRBs at only a factor of $\sim 10\times$ to $25\times$ fainter than our faintest burst, possibly representing the Galactic counterpart of an FRB. Continuing to probe the faint end of the FRB energy 
distribution could bridge the gap between the brightest Galactic bursts and the faintest FRBs. 

\subsection{Resolving aliasing ambiguities in the 16.35-day periodicity }
\frb\ stays in the CHIME/FRB field of view for tens of minutes every sidereal day. The beating of the CHIME/FRB regular exposure pattern with the intrinsic 16.35-day period leads to a degeneracy between frequency $f_0=(P_0)^{-1}=(16.35~\rm{day})^{-1}$ and an alias $f_N = N\,f_\mathrm{sid} \pm f_0$, where $N$ is an integer, 
$f_\mathrm{sid} = (0.99727~\rm{day})^{-1}$ is the frequency of a sidereal day \citep{chime2020a}. An upper limit on $N$ can be estimated from the observed duty cycle over exposure time. With 5/16.35-day fractional active phase and 12~min exposure per sidereal day ($\sim 0.008$ sidereal day), $N_\mathrm{max}=5/16.35/0.008=38$. 

\begin{figure}
    \includegraphics[width=1.0\linewidth,scale=1.0]{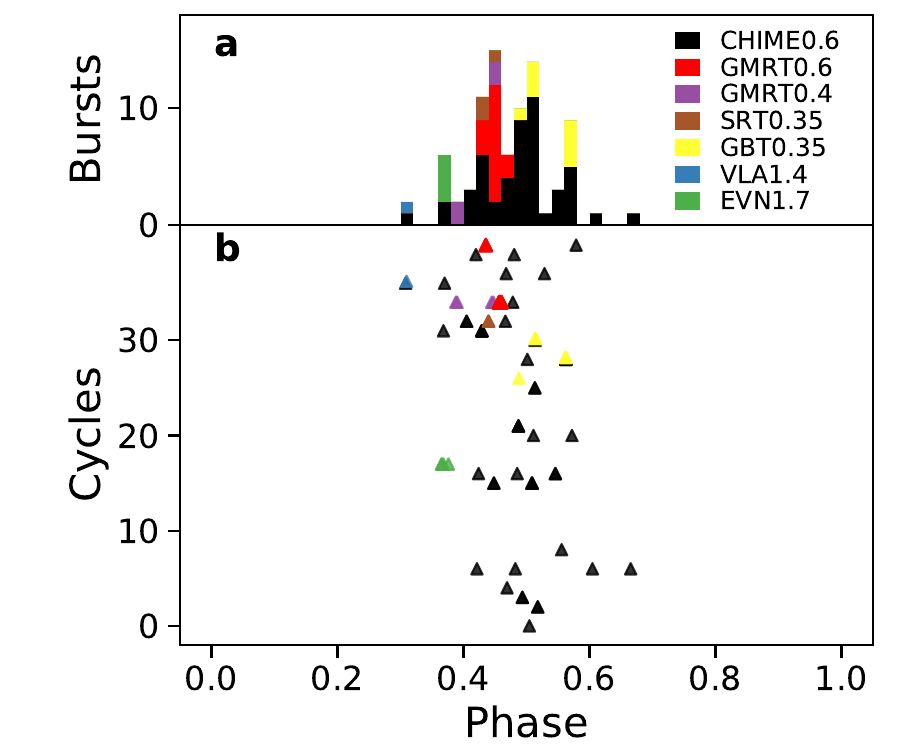}
    \vspace{-5mm}
    \includegraphics[width=1.0\linewidth,scale=1.0]{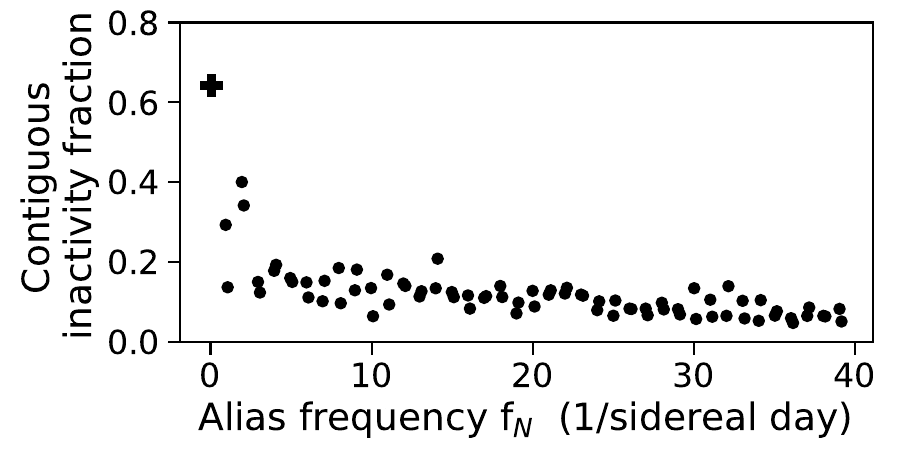}
        \caption{\emph{Top:} ToA of bursts from \frb\ folded at a period of 16.35 days with MJD 58369.18 referenced as phase 0. Bursts captured by different facilities \citep{mnh+20,chime2020a,20Pilia,20Chawla,20Aggarwal,20ATelGMRT} are shown in different colours as specified in the legend. The numbers in the legend are the central frequency of the observation in unit of GHz. (a) Number of bursts at each phase. (b) The phase of detection for individual cycles. 
        \emph{Bottom:} The longest contiguous phase region without detectable activity for the plausible beat periods. The dark cross is for the 16.35-day periodicity.
                }
        \vspace{-5mm}

    \label{fig:fold}
\end{figure}

These aliasing ambiguities can be resolved using burst detections at other observatories, as they can observe during the inactive windows of the aliased solutions. Using the bursts detected by CHIME/FRB and all published bursts from other observatories, 
including the 15 described in this work,
we fold at different alias periods. As shown in Figure~\ref{fig:fold}, the contiguous inactive fraction  \citep{20Rajwade} is much smaller for all beating periods, as compared to the 16.35-day period. This indicates that the bursts are falling outside of the active window for the alias frequency, disfavouring the possibility of an aliased period being the real period. 

\subsection{Imaging}
\label{sec:imaging}
Using the interferometric visibilities that 
were simultaneously recorded, we can localize the brightest bursts and also search for any persistent radio emission associated with the FRB. The data of 2020 March 24 were processed and calibrated using the CASA software package \citep{mws+07}. We used 0217+738 as a complex gain calibrator. 

\begin{figure*}
     \includegraphics[width=0.329\linewidth,scale=1.0]{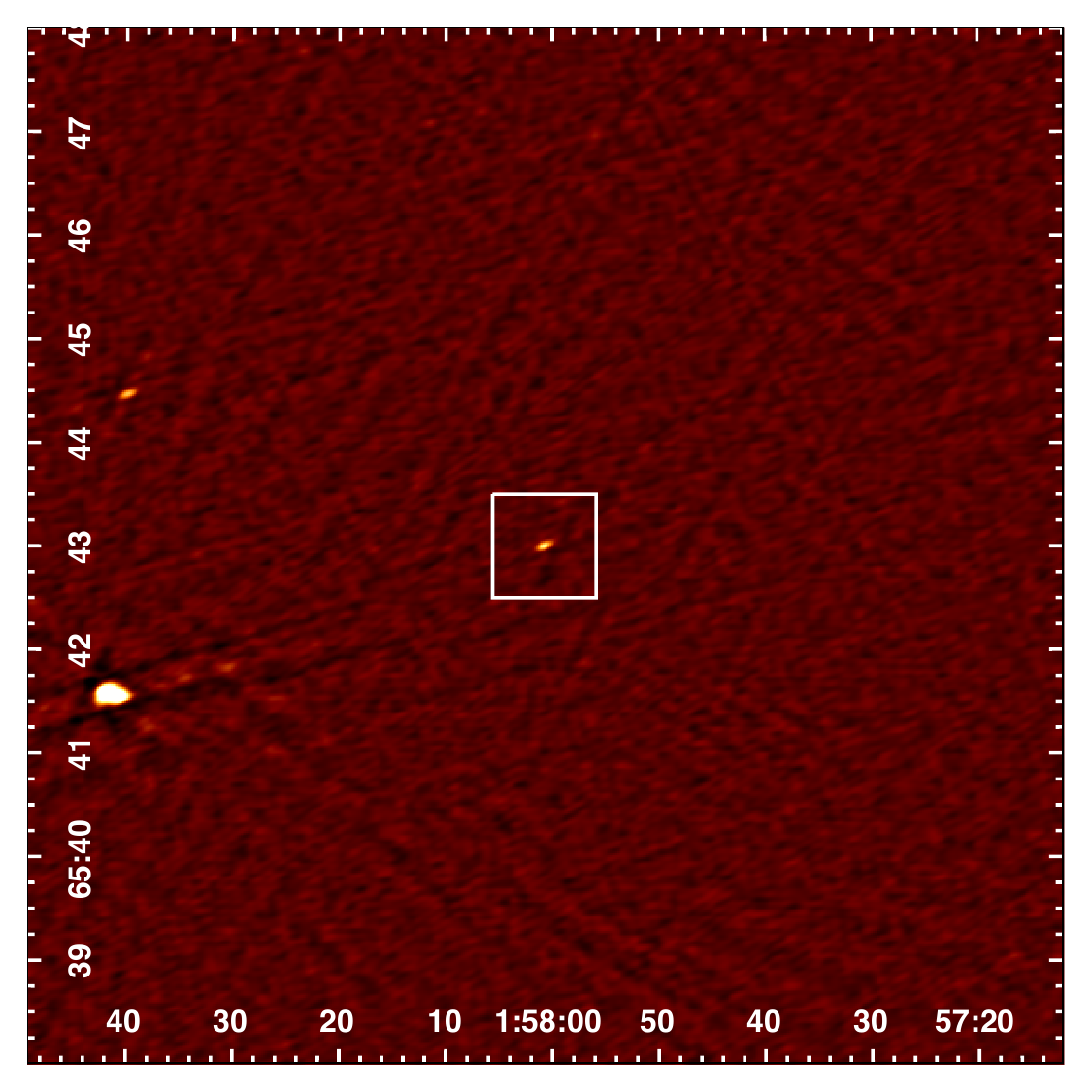}%
     \begin{picture}(0,0)
     \put(-66,100){\includegraphics[width=0.125\linewidth,scale=1.0]{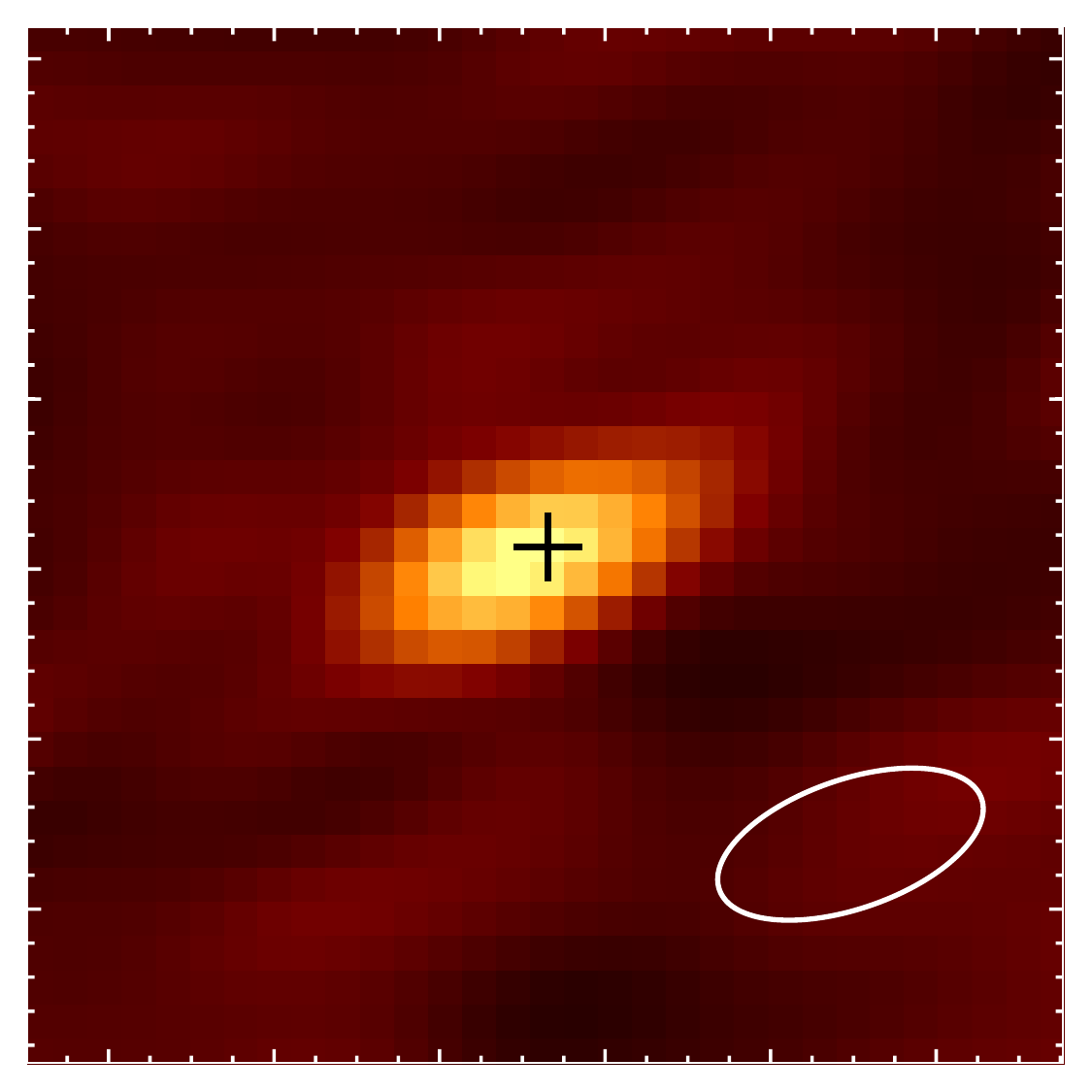}}
     \end{picture}
     \includegraphics[width=0.329\linewidth,scale=1.0]{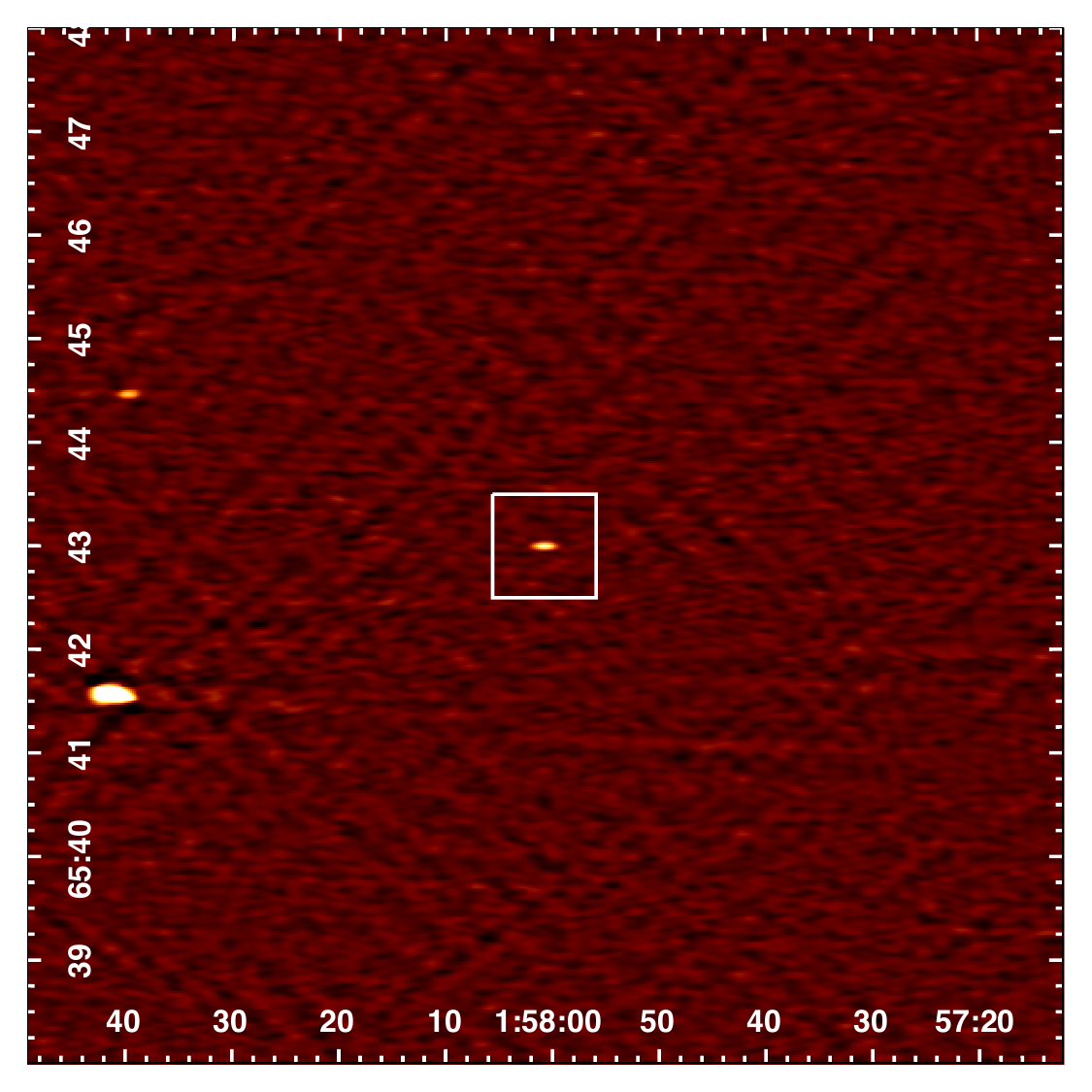}%
     \begin{picture}(0,0)
     \put(-66,100){\includegraphics[width=0.125\linewidth,scale=1.0]{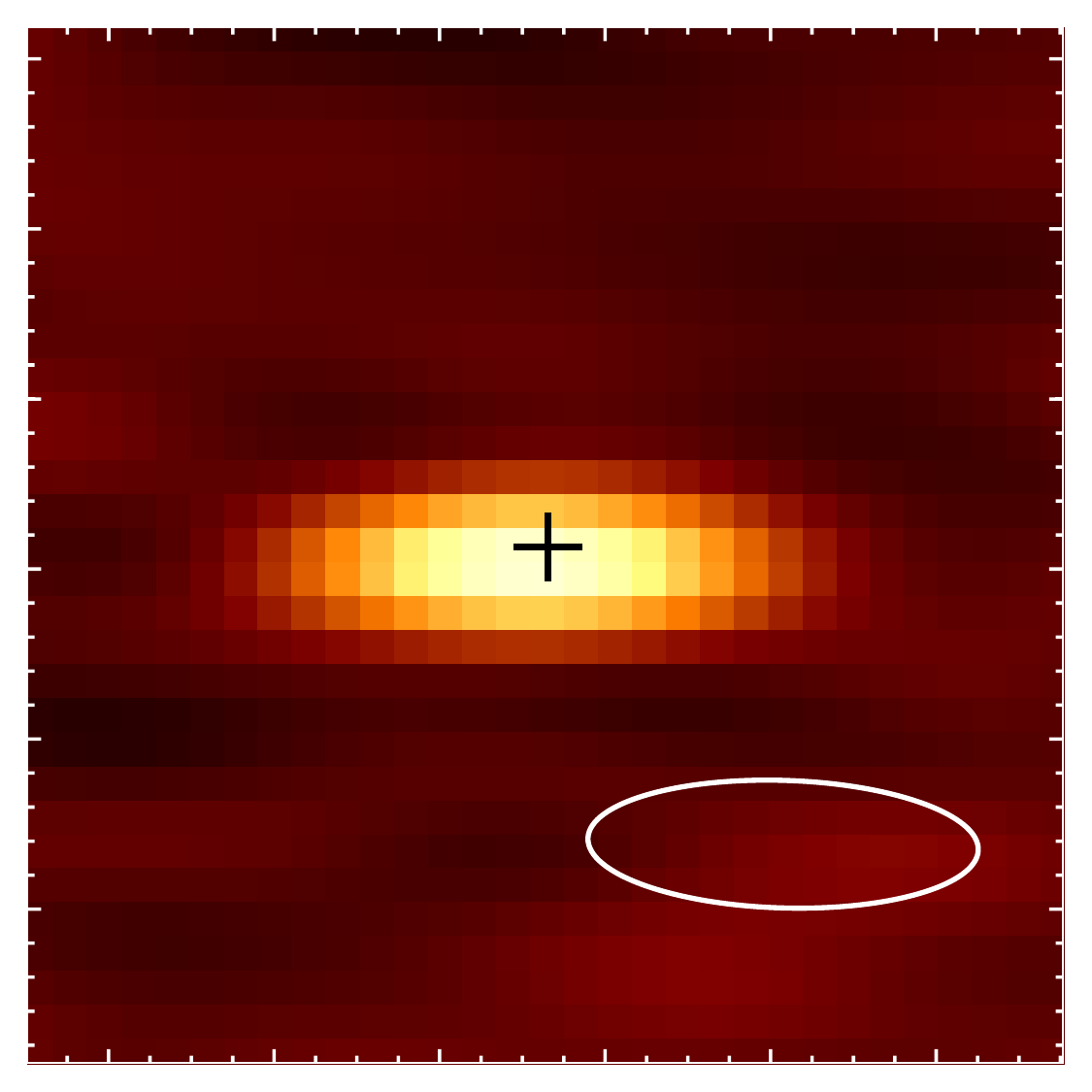}}
     \end{picture}
     \includegraphics[width=0.329\linewidth,scale=1.0]{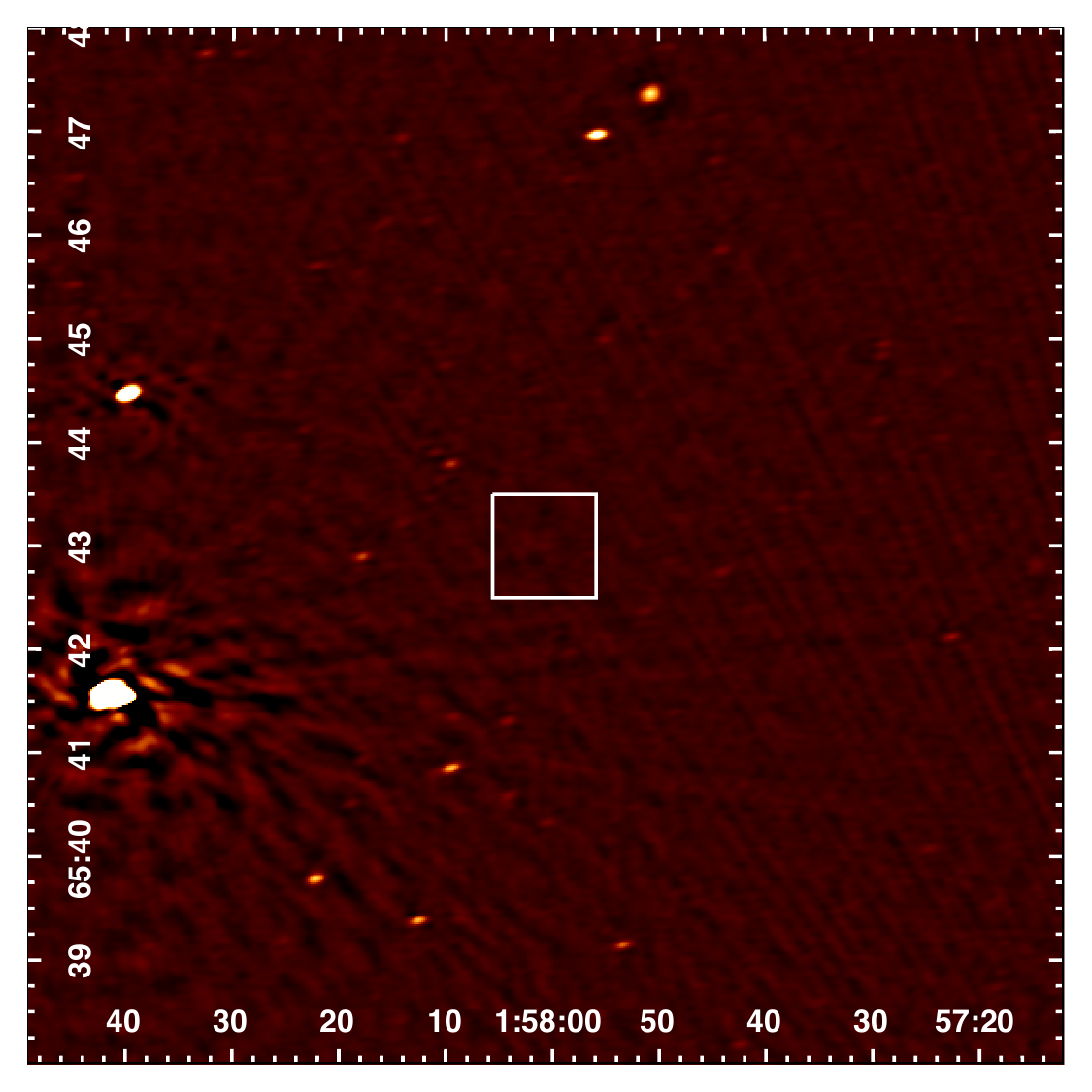}
     \caption{Images of $10\arcmin~\times~10\arcmin$ region centered 
              on the position of \frb, which is indicated by 
              $1\arcmin \times 1\arcmin$ box.  The left (Burst~04) and 
              middle (Burst~11) plots show the detection of single bursts using 
              visibilities from only two (Burst~04) or one (Burst~11) 2.68~sec 
              time samples.  The inset of each plot shows the $1\arcmin \times 1\arcmin$
              region around the burst.  The CLEAN beam is shown in the bottom right 
              of each inset and the EVN position is indicated by a black cross.
              The right plot shows a deep image made using data from the full 
              $\approx2$~hr observation of 2020 Mar 24.
              }
     \label{fig:burstimage2}
\end{figure*}

Even though the time resolution of the visibility 
data was 2.68~sec, two of the bursts from 24~March 
were bright enough for an imaging detection.  Since 
the dispersive delay across the band is only 
$\approx 2.2$~sec, we did not attempt to de-disperse 
the visibility data.  For Burst~04 we made an image from 
two time samples and for Burst~11 we made an image from 
one time sample of the calibrated visibility data.  
For each image, we also did one round of 
phase-only self-calibration on other bright sources in the field.
Burst~04 was detected with a $\rm S/N = 21$ and Burst~11 
with a $\rm S/N =27$.  Both bursts could be localized to 
a $5\sigma$ positional uncertainty of $\lesssim 1\arcsec$.
Although a precise localization has already been determined
with VLBI for this source \citep{mnh+20}, our results 
provide an interesting proof-of-concept for future 
sub-arcsecond localizations with the GMRT. 
The instantaneous images of bursts 04 and 11 are shown 
in Figure~\ref{fig:burstimage2}.

No source was detected at the FRB position in the image (see Figure~\ref{fig:burstimage2})
made with the full $\sim$2~hr observation on 24~March, which 
allows us to set a $3\sigma$ upper limit on the flux density 
at 650~MHz of $66~\mu \rm Jy$.  \citet{mnh+20} used the VLA 
at 1.6~GHz to set a $3\sigma$ flux density limit of 
$18~{\mu \rm Jy}$, which is more constraining than our 
result for spectral indices $\alpha > -1.4$.  

\subsection{Periodicity search on small timescales}
We performed a Fourier domain acceleration search using \PRESTO’s \texttt{accelsearch} algorithm on the $\sim 2$\,hr on-target data on 2020 March 24, in the DM range of $348.82 \pm 5\ \mathrm{pc\,cm}^{-3}$ with periods between 1\,ms to 20s. The full data search was sensitive to accelerations up to $\sim$28\,ms$^{-2}$ for a periodicity of 10\,ms, while 20-minute segments of the data were also searched making us sensitive to accelerations up to $\sim$200\,ms$^{-2}$. In addition, we performed a jerk search \citep{Andersen+2018} to account for changing accelerations within the span of our observation. 

A number of weak candidates have shown up in the search, including a 15.6~ms ($\sim 6 \sigma$) periodicity candidate in one of the 20-minute data segments. However, this periodicity is not detected with any significance when we search the full $\sim$2\,hr observation.

\section{SUMMARY AND CONCLUSIONS}\label{sec:discussion}
We have outlined the uGMRT detection of 15 bursts, of which 12 is the largest number of single-session detections, including the faintest ones to date. The burst rate appears to be highly variable and the faintest burst is only $\sim10\times$ brighter than the brightest Galactic burst detected till date.

We have effectively resolved aliasing ambiguities in the 16.35-day period. A survey telescope like CHIME is well tuned to discovering new FRBs, while monitoring and follow-up observations with a steerable radio telescope like the GMRT can break the degeneracy between the intrinsic period and an alias caused by the regular exposure pattern of CHIME. 

Further sensitive observations such as these will be key to detecting any underlying short-timescale periodicity, as we continue to detect bursts with short separations. We will continue to observe during the active window to investigate the origin of the 16.35-day periodicity. For example, any modulation of the scintillation of the bursts as a function of the phase within the active window is evidence for orbital motion around a companion. Sensitive polarization observations in the future can help in constraining emission models and mechanisms.

We have imaged two of the brightest bursts with sub-arcsecond precision, showing the promise of the GMRT to localize known repeating FRBs. The GMRT can also potentially localize newly detected repeating FRBs, especially since its field of view is well-matched with the CHIME beam. While the time resolution of the uGMRT interferometer is currently limited to 670ms, the increased sensitivity compared to CHIME will allow for imaging of the brighter CHIME bursts, above $\gtrsim 3\,$\flu.

\section{ACKNOWLEDGEMENTS}
We thank the staff of the GMRT who have made these observations possible. GMRT is run by the National Centre for Radio Astrophysics of the Tata Institute of Fundamental Research. We acknowledge use of the CHIME/FRB Public Database, provided at \url{https://www.chime-frb.ca/} by the CHIME/FRB Collaboration. We acknowledge the support of the Natural Sciences and Engineering Research Council of Canada (NSERC) (funding reference number RGPIN-2019-067, CRD 523638-201). We receive support from Ontario Research Fund - Research Excellence Program (ORF-RE), Canadian Institute for Advanced Research (CIFAR), Canadian Foundation for Innovation (CFI), Simons Foundation, Thoth Technology Inc. and Alexander von Humboldt Foundation. RSW acknowledges financial support by the European Research Council (ERC) for the ERC Synergy Grant BlackHoleCam under contract no. 610058.

\section{DATA AVAILABILITY}
The data underlying this article will be shared on reasonable request to the corresponding authors.

\bibliographystyle{mnras}
\bibliography{R3}

\end{document}